\documentclass{PoS}

\title{Recent Results from The Askaryan Radio Array}

\ShortTitle{Recent Results from The Askaryan Radio Array}

\author{ \speaker{Amy Connolly},  for the ARA Collaboration\footnote{for collaboration list see PoS(ICRC2019)1177.  ARA author list available at http://ara.wipac.wisc.edu/authorlist.}
 \\
        The Ohio State University\\
        E-mail: \email{connolly@physics.osu.edu}
        }

\abstract{The Askaryan Radio Array (ARA) is an ultra-high energy (UHE) neutrino
telescope at the South Pole consisting of an array of radio antennas 
aimed at detecting the Askaryan radiation produced by neutrino interactions 
in the ice.    Currently, the experiment has five stations in operation that have been deployed in stages since 2012. This contribution focuses on the development of a search for a diffuse flux of neutrinos in two ARA stations  (A2 and A3) from 2013-2016.
A background of $\sim 0.01-0.02$
events is expected in one station in each of  
two search channels in horizontal- and vertical-polarizations.  The 
expected
new constraints on the flux of ultra-high energy neutrinos based on four years of analysis with two stations
improve on the previous limits set by ARA by a factor of about two.  The projected sensitivity
of ARA's five-station dataset is beginning to be competitive with other neutrino telescopes at high energies 
near $10^{10.5}\,$GeV.
}

\FullConference{36th International Cosmic Ray Conference -ICRC2019-\\
		July 24th - August 1st, 2019\\
		Madison, WI, U.S.A.
}

\begin{document}

\section{Introduction}

Neutrinos at extreme energies 
($>10$\,PeV) are unique particle 
messengers to the distant, high energy universe. 
As chargeless and only weakly interacting particles, unlike cosmic rays and gamma rays, neutrinos can arrive from cosmic distances 
unattenuated and point back to their sources. High-energy neutrinos are expected to have two components.  The ``astrophysical'' flux is produced by the cosmic ray accelerators themselves, and is being observed by IceCube at energies below
ARA's threshold.
There is also an expected  ``cosmogenic'' neutrino flux in ARA's energy 
regime from interactions between
cosmic rays and the cosmic microwave background light.

Detection of these UHE neutrinos is challenging because of their low fluxes ($<1$ event/yr/km$^3$) and low cross section ($\sigma \sim 10^{-31}\,\textrm{ cm}^2$),
so that a detector with an instrumented volume approaching $100\,\mathrm{km}^3$ is necessary, which is not feasible using the optical-Cherenkov technology of experiments such as IceCube~\cite{Aartsen:2016nxy} or the Cubic Kilometre Neutrino Telescope (KM3NeT)~\cite{Adrian-Martinez:2016fdl}. Instead, one can utilize the radio Cherenkov (Askaryan) emission produced by the relativistic neutrino-induced showers \cite{Askaryan:1962hbi, Askaryan:1965}. Because the attenuation length of radio waves in glacial ice is of order 1\,km 
~\cite{Barwick:2005zz}, instrumentation can be deployed more sparsely for
a given detection volume relative to the optical technique.  The Askaryan Radio Array (ARA) detector seeks to observe neutrinos above 10 PeV through the Askaryan effect, and is under construction at the South Pole.  This proceeding reviews the latest results from ARA.

\section{The ARA Detector}

\begin{figure}[!tbp]
  \centering
    \begin{minipage}[b]{0.49\textwidth}
    \includegraphics[width=\textwidth]{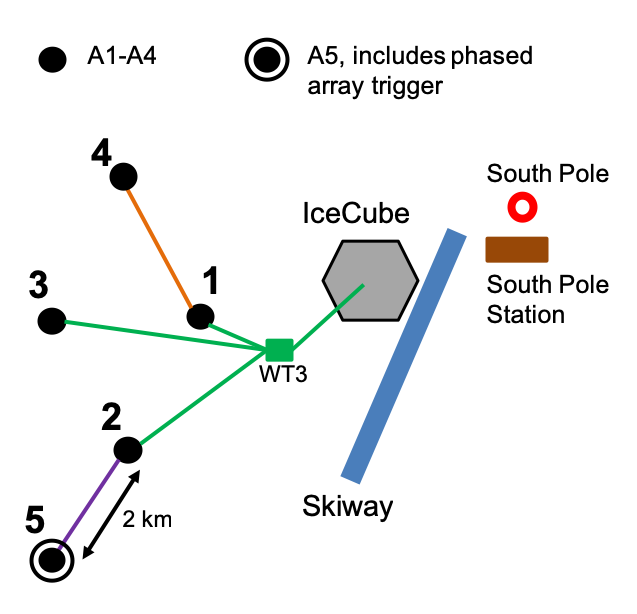}
    \caption{A map of the current five-station ARA instrument at the South Pole.}
    \label{fig:ara_map}
  \end{minipage}
    \hfill
  \begin{minipage}[b]{0.49\textwidth}
    \includegraphics[width=\textwidth]{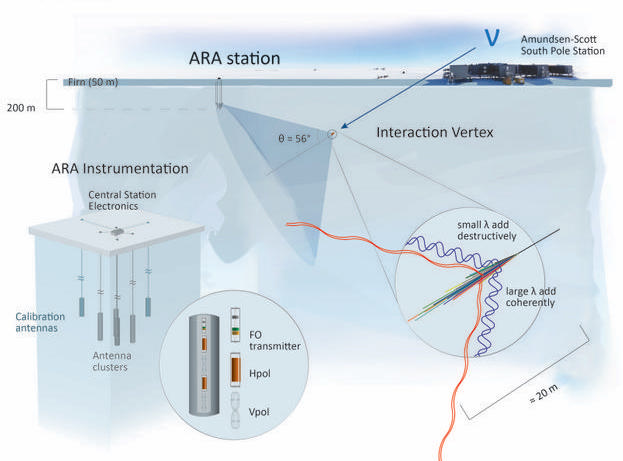}
    \caption{A diagram of an ARA station and neutrino detection concept.}
    \label{fig:ara_station}
  \end{minipage}
\end{figure}

The Askaryan Radio Array consists of five  neutrino-detecting stations buried in the Antarctic ice sheet at the South Pole at up to 200\,m depth,
each operating independently. A map of these five stations and the South Pole Station can be seen in Fig.~\ref{fig:ara_map}. The ARA instrument, beginning with a prototype
``Testbed'' followed by five stations up to 200\,m deep,
was deployed in Austral seasons
from 2011-2012 to 2017-2018, inclusive.
The fifth deep station includes a threshold-lowering phased-array trigger~\cite{Allison:2018ynt}. The full five-station array is denoted ``ARA5,'' with individual stations identified as A1, A2, etc.

Each station is composed of four measurement strings, with each string holding two vertically-polarized (VPol) antennas and two horizontally-polarized (HPol) antennas, near the bottom of the string as shown in Fig.~\ref{fig:ara_station}. The antennas are broadband, with azimuthally-symmetric birdcage dipoles for VPol and ferrite-loaded, quad-slot antennas for HPol. After each antenna, the signal is bandpass-filtered to 150-850\,MHz, and notch-filtered at 450 MHz, the latter being used for radio communications at the South Pole. After filtering, the signal passes through an amplifier, and then is converted to an optical signal and transmitted to the Data Acquisition (DAQ) electronics box, located on the ice surface. The recorded events are stored locally before being transferred via fiber-optic network to a central storage facility at the IceCube Counting Laboratory (ICL). 

In addition to the measurement strings described above, 
calibration pulser strings are deployed approximately 41\,m from each station center. These local ``cal pulsers'' are typically fired once every second, and are used to calibrate the station geometry and the inter-channel timing of the measurement antennas. The cal pulsers also serve as a convenient marker of station health.

An ARA station triggers when in 3/8 antennas of the same polarization, a power envelope of the waveform exceeds 5-6 times the thermal noise level.  The three excursions
need to be coincident within the maximum causal travel time for a light ray across the array, which is about 170~ns for most stations.
The threshold for the single-channel trigger is adjusted in real time by a servo to maintain a 5\,Hz global trigger rate for each station. In A5, the ARA trigger described above is augmented by a threshold-lowering phased-array triggering system, which coherenty adds the signals from different antennas together before triggering. \cite{Allison:2018ynt}.

\section{Search for a Diffuse Flux of Neutrinos in ARA Station 2}

ARA has previously published the analysis of the first 10 months of recorded data of A2 and A3 \cite{Allison:2015eky}. We are nearing the completion of
two parallel searches for a diffuse flux of neutrinos
in the same two stations from 2013 to 2016, inclusive. In this proceeding, we focus on the analysis of data from A2. All voltages and time sampling of waveforms are calibrated using the methods
developed in the previous publication, and details can be found therein~\cite{Allison:2015eky}. The analysis was 
performed ``blind,''  with 10\% of the data set aside in a burn sample used to tune cuts and understand backgrounds, before those cuts are 
imposed on the remaining 90\% of the dataset to
search for any candidate events. This analysis does not address data taken in A1, A4 and A5 as the calibration of those stations is underway.

\subsection{Detector Livetime}
\begin{figure}
    \centering
    \includegraphics[width=\linewidth]{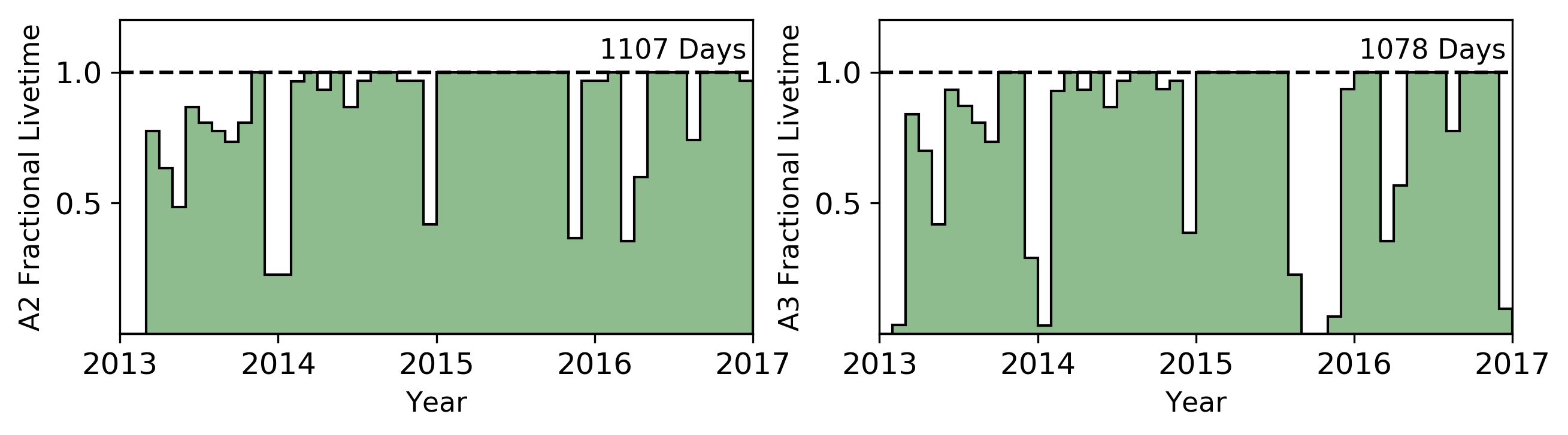}
    \caption{Histograms showing the fractional livetimes for A2 (left) and A3 (right) from deployment in 2013 through the end of the analysis period in 2016. Only about 3 years of data are analyzable, from the 4 years of deployment.}
    \label{fig:A23_uptime}
\end{figure}

Fig.~\ref{fig:A23_uptime} shows the fractional livetime of A2 and A3 over the four year analysis period.  Both 
stations have accumulated approximately 1100 days of 
livetime since deployment. 
In analyses, we exclude runs that contain calibration activities (e.g., operation of a surface pulsing unit or the IceCube deep pulser) and remove livetime that is contaminated with anthropogenic activity. These cuts keep $\sim98$\% of our livetime in A2 for analysis, compared to the 62\% achieved in the Testbed searches. In analyses, the A2 livetime is split up into five configurations that represent different run conditions such as readout length and trigger settings.

\subsection{Simulation}
AraSim~\cite{Allison:2014kha} is the simulation used by 
the ARA experiment used to model neutrino and noise events  to assess neutrino efficiencies.  For this search,
AraSim generates neutrinos from a diffuse flux 
in a cylindrical fiducial volume, and 
 Askaryan emission using a parametrized shower~\cite{2011PhRvD..84j3003A}. AraSim propagates the signal taking into account signal attenuation and  ray bending in the ice based on a depth-dependent index of refraction model. 
Signals are convolved with the antenna responses and subsequently the electronics (filter and amplifier) responses are applied. Lastly, the convolved waveforms
are
subject to the triggering algorithm. If the event triggers, it is stored in a manner that mimics that of real data so that our analysis code can take either data or simulated events interchangeably.

\subsection{Cuts}


This analysis imposes an event filter to reduce the size of the data set in advance of computationally 
expensive analysis. At 5\,Hz, each ARA station records approximately $10^{8}$ events per year. 
To reject events that triggered on thermal noise
fluctuations, we apply fast ($60$\,ms/event) thermal noise rejection algorithms described in previous proceedings~\cite{Lu:2017amt}. These algorithms reduce the data quantity by over an order of magnitude while being  approximately 90\% efficient for neutrino
events from $10^{8}$ to $10^{12}$\,GeV.

Events that are contaminated by continuous wave (CW)
noise are either removed from the analysis
or filtered of their contamination before advancing to the next stage of the analysis.
CW interference is characterized by a strong spectral peak at one or a few frequencies within a narrow band. The most common type of CW encountered in ARA is generated by the $\sim$403\,MHz radiosonde attached to weather balloons that are launched twice daily from the South Pole.  One such filter notches spectral amplitudes
around contaminating frequencies, and adjusts the phases within the notch. This filter was used in an ANITA analysis, and adapted for use in ARA~\cite{Allison:2018cxu}.

After filtering of CW interference, we perform an interferometric reconstruction on each event to determine the direction of the source of the measured signals~\cite{Romero-Wolf:2014pua}. For every point on the sky, characterized by an elevation and azimuthal angle $(\theta, \phi)$, we can find the expected delay $\tau(\theta, \phi)$ between time of arrival of signals for any pair of antennas. The calculation of the delay takes into account the depth-dependent index of refraction of Antarctic ice. After adjusting for this delay, we compute the cross-correlation value $C_{i,j}$ between every pair of antennas $i$ and $j$, and then sum the correlation values for all pairs for that point on the sky, $C_{sky}$. The peak in the summed correlation map is taken as the source direction. We reject an event if it reconstructs to above the surface, as we are searching for neutrino interactions in the ice. 

\begin{figure}
    \centering
    \includegraphics[width=\linewidth]{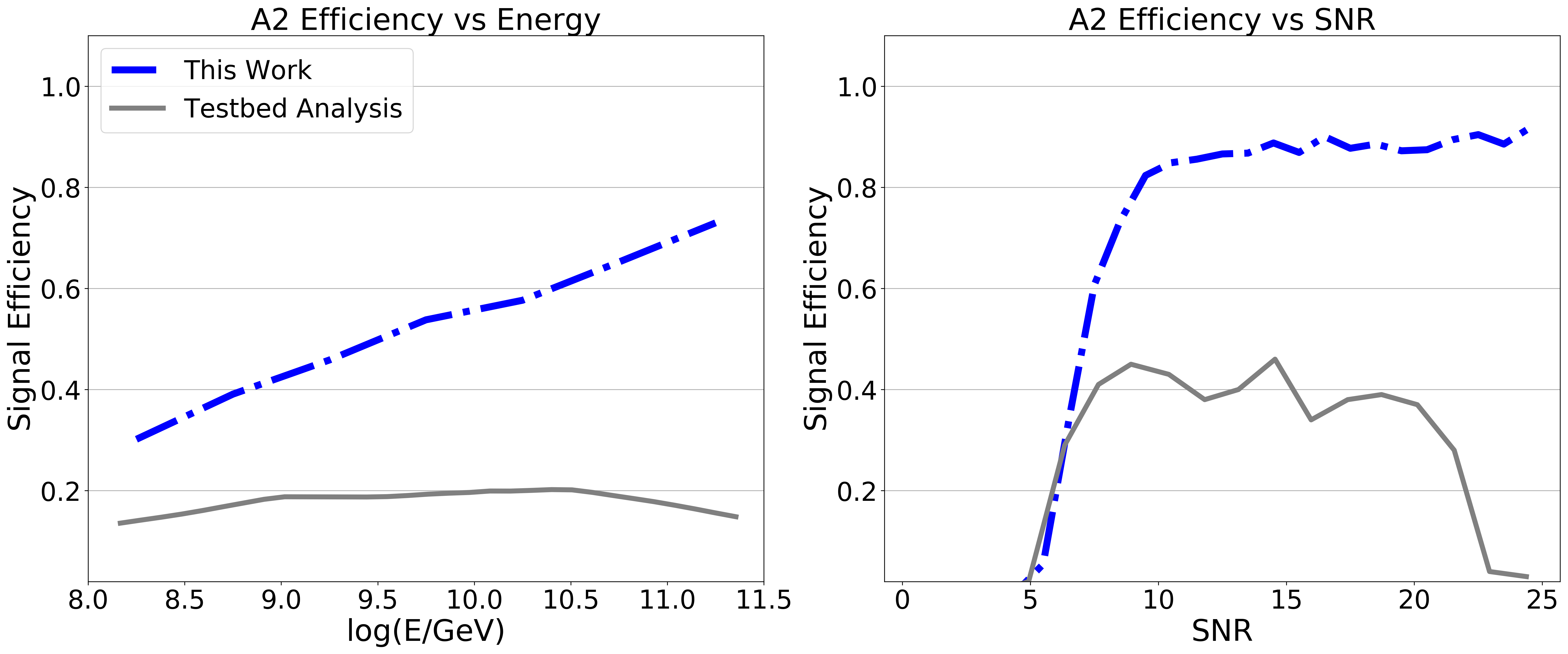}
    \caption{The signal efficiency of the A2 analysis compared to that of the previous Testbed analysis~\cite{Allison:2014kha} as a function of the neutrino energy (left) and as a function of  signal-to-noise ratio (right).}
    \label{fig:A2_efficiency}
\end{figure}

We place a final cut in a two-dimensional parameter space defined by the signal-to-noise ratio (SNR) and the summed correlation value ($C_{sky}$) discussed above. The SNR is defined as the third-highest peak voltage in an event divided by the average RMS noise level in that channel. The final cuts are optimized for the best expected 
limit assuming an observation will come from backgrounds alone in the 90\% data sample. 

\subsection{Efficiencies and Expected Background}

Fig.~\ref{fig:A2_efficiency} shows the analysis efficiencies for neutrinos in this search. On the left is the signal efficiency as a function of energy, and on the right is the signal efficiency as a function of SNR. 
At a given energy, we achieve an efficiency of approximately 35\% at the lowest energies near $10^{8}$\,GeV, and a maximum efficiency of nearly 75\% at the highest energies near $10^{12}$\,GeV. As a function of SNR, our efficiency begins to turn on at an SNR of 5, reaches its 50\% efficiency point near an SNR Of 8, and saturates at about 90\% by an SNR of 10. For comparison, we show the efficiency of the previous Testbed analysis \cite{Allison:2014kha}, which is the previous ARA
analysis that is the most comparable to the one presented here. This analysis achieves a factor of 2-4 improvement in efficiency across the energy range relative to the Testbed analysis, and plateaus at about a factor of two higher than that analysis at high SNR.

\begin{figure}
    \centering
    \includegraphics[width=0.8\linewidth]{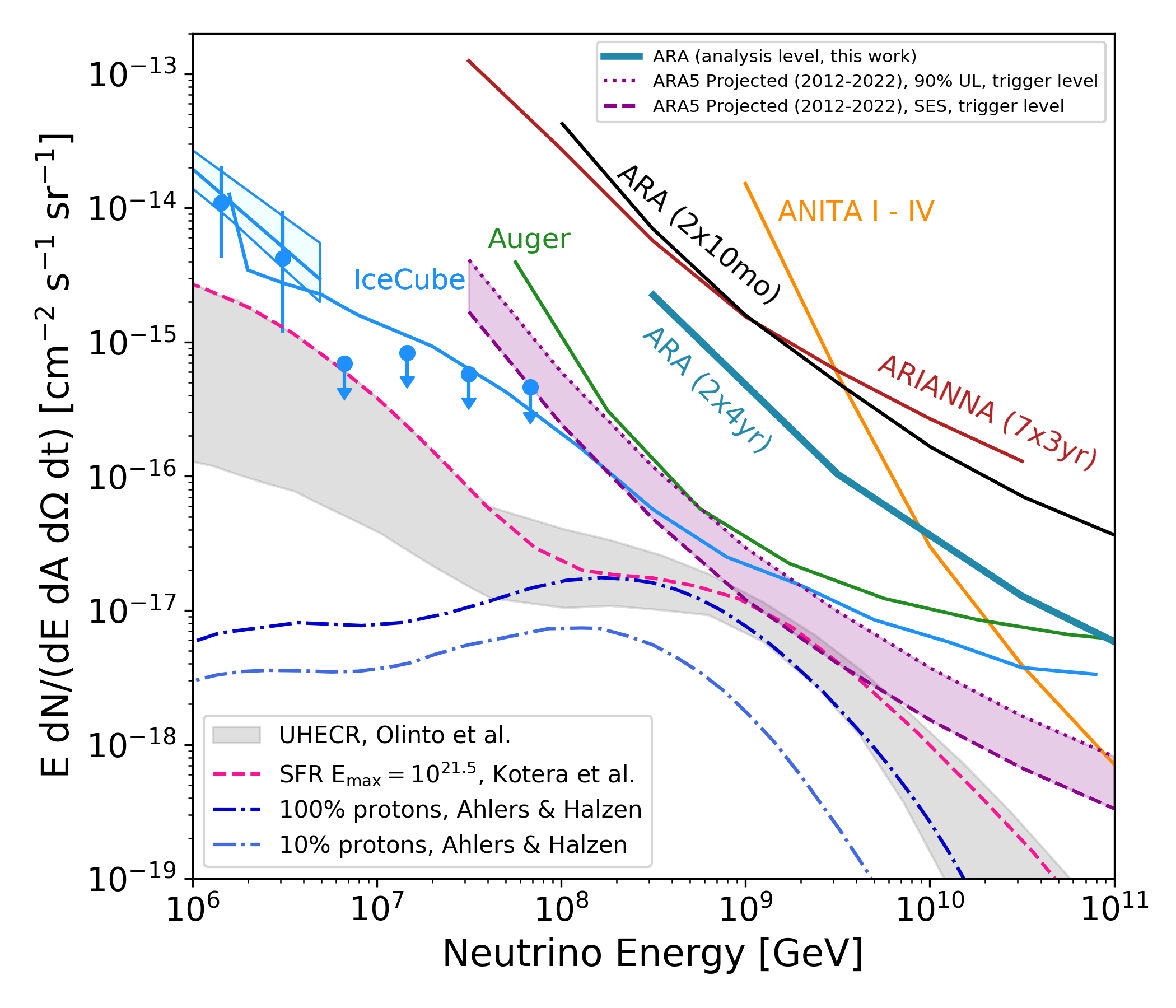}
    \caption{The 90\% CL upper limit, including analysis efficiencies, expected for this work including A2 and A3 (thick aqua curve); also shown in the purple band is the projected trigger-level sensitivity of the full five-station array at the 90\% CL and at the single event sensitivity (SES) level. For comparison, we also show the latest results from IceCube \cite{Aartsen:2016xlq, Aartsen:2018vtx}, Pierre Auger \cite{Aab:2015kma}, ANITA \cite{Gorham:2019guw}, and ARIANNA \cite{PersichilliThesis}; cosmogenic neutrino flux predictions are from Olinto \textit{et. al.} \cite{Olinto:2011ng}, Kotera \textit{et. al.} \cite{Kotera:2010yn}, and Ahlers \& Halzen \cite{Ahlers:2012rz}.}
    \label{fig:limits}
\end{figure}

This analysis was performed in two search channels
for each polarization, VPol and HPol. We estimate
the backgrounds by extrapolating from a model of
the distribution of background events designed on
the 10\% burn sample. 
We expect  $0.010^{+0.003}_{-0.004}$ background events in VPol and $0.016^{+0.003}_{-0.003}$ in HPol in one of the analyses in A2.

\subsection{Expected Limit}

In Fig.~\ref{fig:limits}, we show the expected 90\% CL upper limit from the four year analysis of data from A2 \& A3 
if the number
of events observed
in the 90\% data samples are consistent with the background
expectations.  The expected limit from this analysis is presented in the solid aqua curve, and is plotted in the context of other experimental efforts as well as predictions for the cosmogenic neutrino flux. In the expected limit, we assume that A3 has the same sensitivity as A2, as is expected.

We also show in a purple band the projected trigger-level sensitivity of the full five-station station array (ARA5) by 2022.  The upper edge of the band is the 90\% confidence-level upper limit assuming an observation consistent with background; the lower edge of the band is the single-event sensitivity. In this projection, we take into account the additional 9 station-years of livetime that is already archived but not yet analyzed, in addition to 17.5 station-years of livetime that will be accumulated by 2022, which means the total livetime in the five-station study is approximately 34.5 station years (8+9+17.5), or more than a factor of four times the 8 station-years
presented here. We include the effect of the threshold-lowering phased array on A5, which increases the effective volume of that station by 20-60\% depending on energy \cite{Allison:2018ynt}. By 2022, at the trigger level, we expect a factor of approximately five improvement in sensitivity over the analysis presented here. As can be seen, ARA5 will have accumulated sufficient data to set world-leading limits above $\sim10^{10}$\,GeV within the next three years, or observe the first ultra-high energy neutrino candidates.

\section{Summary and Outlook}
In this proceeding, we have discussed the analysis of data taken from 2013-2016 in ARA station A2, and presented an expected limit on stations 
A2 \& A3. The analysis is a factor of 2-4 more efficient than previously achieved in the initial analysis of Testbed data, 
and keeps approximately 98\% of the recorded detector livetime for analysis. 

With the current five stations, one of which includes a threshold lowering phased-array trigger, ARA is expected to set the world's strongest limit on the flux of UHE neutrinos above $10^{10}$\,GeV by 2022, or observe
the first ultra-high energy neutrinos. Importantly, ARA will constrain several cosmogenic neutrino flux models, and will probe whether the astrophysical neutrino flux measured by IceCube continues unbroken to the ultra-high energy regime. Such path finding will be essential for setting the scale for the next generation of neutrino observatories, e.g. RNO, IceCube-Gen2, and GRAND.

\section{Acknowledgements}
We thank the National Science Foundation Office of Polar Programs and Physics Division for their generous support through Grant NSF OPP-902483, Grant NSF OPP-1359535 and
Grant NSF OPP-1404212. Furthermore, we are grateful to the Raytheon Polar Services Corporation and the Antarctic Support Contractor for field support. A. Connolly thanks the National Science Foundation for their support through NSF award 1806923, NSF CAREER award 1255557, BIGDATA Grant 1250720, and thanks the United States-Israel Binational Science Foundation for their support through Grant 2012077. A. Connolly is also grateful to the Ohio Supercomputer Center (OSC) for computational resources.

\bibliographystyle{JHEP} 
\bibliography{references}

\end{document}